\documentclass[11pt]{article}
\usepackage{a4wide}

\usepackage{graphics}
\usepackage{fullpage,epsf,graphicx, amstext,url} 
\usepackage{bm}
\usepackage{capt-of}
\usepackage{fancyhdr}
\usepackage{amsmath,amssymb,mathtools,physics,braket}
\usepackage{rotating}
\usepackage{float}
\usepackage{epstopdf}
\usepackage{indentfirst}
\usepackage{makecell}
\usepackage{textcomp,gensymb}
\usepackage[colorlinks=true,citecolor=blue]{hyperref}
\usepackage[toc,page]{appendix}
\usepackage{mathrsfs}
\usepackage{calligra}
\DeclareMathAlphabet{\mathcalligra}{T1}{calligra}{m}{n}

\usepackage[style=ieee, citestyle=numeric-comp, backend=bibtex]{biblatex}
\usepackage[compat=1.1.0]{tikz-feynman}

\usetikzlibrary{calc,patterns,angles,quotes,math,decorations.text,arrows.meta,decorations.markings}
\tikzset{->-/.style={decoration={
  markings,
  mark=at position #1 with {\arrow{stealth}}},postaction={decorate}}}

\usepackage{bbm}

\renewcommand{\[}{\left[}
\renewcommand{\]}{\right]}
\renewcommand{\(}{\left(}
\renewcommand{\)}{\right)}
\newcommand{\nn}{\nonumber}
\newcommand{\dg}{\dagger}
\newcommand{\Mpl}{M_{{Pl}}}
\newcommand{\calO}{\mathcal{O}}
\newcommand{\calS}{\mathcal{S}}
\newcommand{\calU}{\mathcal{U}}

\newcommand{\calC}{\mathcal{C}}
\newcommand{\calH}{\mathcal{H}}

\newcommand{\calA}{\mathcal{A}}
\newcommand{\calB}{\mathcal{B}}
\newcommand{\bx}{\mathbf{x}}
\newcommand{\bk}{\mathbf{k}}
\newcommand{\bp}{\mathbf{p}}

\newcommand{\Hilbert}{\mathscr{H}}

\DeclareMathOperator{\arcsinh}{arcsinh}

\begin{document}

{\renewcommand{\thefootnote}{\fnsymbol{footnote}}
		
\begin{center}
{\LARGE Momentum-space entanglement entropy in de Sitter} 
\vspace{1.5em}

Suddhasattwa Brahma$^{1}$\footnote{e-mail address: {\tt suddhasattwa.brahma@gmail.com}}, 
Jaime Calder\'on-Figueroa$^{1,2}$\footnote{e-mail address: {\tt jaime.calderon@ed.ac.uk}},
Moatasem Hassan$^{1}$\footnote{e-mail address: {\tt m.k.h.hassan@sms.ed.ac.uk}},
and 
Xuan Mi$^{1}$\footnote{e-mail address: {\tt x.mi-3@sms.ed.ac.uk}}
\\
\vspace{1.5em}

$^{1}$Higgs Centre for Theoretical Physics, School of Physics and Astronomy,\\ University of Edinburgh, Edinburgh EH9 3FD, UK\\[2mm]

$^2$Departamento de F\'isica, Escuela Polit\'ecnica Nacional, Ladr\'on de Guevara E11-253, 170525, Quito, Ecuador\\[2mm]

\vspace{1.5em}
\end{center}
}
	
\setcounter{footnote}{0}

\begin{abstract}
 \noindent We study the momentum-space entanglement between the sub- and super-Hubble modes of a spectator scalar field, with a cubic $\lambda \phi^3$ interaction, in de Sitter space. Momentum-space entanglement has some universal properties for any interacting quantum field theory, and we examine them for this specific curved background using the Hubble scale as a natural delimiter to define UV/IR separation. We show that there are several new subtleties when generalising flat space results due to having a time-dependent interaction term and a non-trivial vacuum state. Our main finding is that the momentum-space entanglement entropy in de Sitter space grows very rapidly, supporting previous similar results for cosmological perturbations \cite{Brahma:2020zpk}, which leads to interesting new questions.
\end{abstract}

\section{Introduction}
de Sitter (dS) space occupies a critical role in our understanding of the cosmos. Both the very early universe as well as the current cosmic epoch can be modelled by dS (or quasi-dS) spacetimes. However, the UV and IR regimes of QFTs in dS suffer from a variety of conceptual issues. For instance, UV physics on Planck scales can lead to an arbitrariness in the choice of the dS vacuum (the so-called `trans-Planckian problem') \cite{Martin:2000xs,Danielsson:2002kx}, unless one assumes that standard Minkowski space underlies physics on Planck scales, not to mention the more drastic claim about the impossibility of obtaining dS space from UV-complete theories of gravity \cite{Vafa:2005ui,Obied:2018sgi,Garg:2018reu,Danielsson:2018ztv, Dasgupta:2018rtp}. On the other end, it is well-known that late-time effects in dS lead to secular divergences and require non-perturbative physics for their resolution \cite{Giddings:2010nc, Polyakov:2012uc,Burgess:2009bs,Burgess:2015ajz}. UV-IR mixing happens naturally due to the non-conservation of energy and red-shifting of the physical degrees of freedom (dofs) from the UV to the IR due to the accelerated expansion of the universe. This makes application of standard Wilsonian renormalisation more tricky in dS. Clearly, there are many aspects of quantum fields in curved spacetimes, and particularly in dS, that are much less understood than their flat-space counterparts. 

Given these considerations, entanglement has become a fruitful avenue to explore these ideas, and is now far from just being an esoteric (albeit, defining) property of quantum fields, having turned into a well-tested physical phenomenon. For example, in the context of gravity, entanglement entropy has been extensively studied to probe the UV structure of spacetime, leading to remarkable insights. Indeed, holography has allowed us to study the entanglement structure of the vacuum by looking at the entanglement associated with a geometric region of space. However, if one wants to explore this through the lens of interacting QFTs, there is a plethora of technical problems to overcome when dealing with position-space entanglement, especially in presence of interacting fields. Nonetheless, entanglement emerges in any subsystem where a sub-algebra of observables can be defined, and this need not be a sub-region of position space. In particular, as shown in \cite{Balasubramanian:2011wt,Lundgren:2019ehi,Kumar:2017ctm}, there is an entanglement entropy associated with different bands in momentum-space. The Fock vacuum being completely factorised for a free theory, all the entanglement in this case comes indeed from interactions. 

Hence, it is meaningful to ask what is the momentum-space entanglement entropy for a quantum field in dS space, looking for deviations from the flat-space results\footnote{See, for instance, \cite{Maldacena:2012xp, 2014JHEP...07..072K, Bhattacharya:2020bal, Bhattacharya:2022wpe, Martin:2021qkg, Boyanovsky:2018soy, Belfiglio:2022yvs, Boyanovsky:2018fxl, Colas:2022kfu} for some recent results on other measures of entanglement in dS space.}. Our main motivation remains cosmology, where the statistics (\textit{e.g.,} the power spectrum, bispectrum, and so on) of the cosmological perturbations, in terms of momentum modes, are typically observed. Evidently, inflation is the most natural test bed to explore these ideas. According to this paradigm, quantum fluctuations are the fundamental seeds from which we can extract the distribution of matter in our universe, and since General Relativity is nonlinear, this implies that such quantum modes must have non-zero entanglement entropy in momentum-space. In order to simplify the technical details of the calculation, and to highlight the salient features associated with entanglement profile of the curved vacuum, we will focus on a spectator scalar field in a pure dS spacetime, the former being a proxy for cosmological perturbations while the latter for inflationary expansion.

A first evaluation for the entanglement entropy of scalar perturbations during inflation, originating from cubic non-Gaussianities, was carried out in \cite{Brahma:2020zpk} (see also \cite{Brahma:2021mng,Burgess:2022nwu}). However, it was realised in that work itself that partitioning the full Hilbert space into sub- and super-Hubble modes would lead to several subtleties over the flat space case. The first one has to do with the ``system'' (super-Hubble) and ``environment'' (sub-Hubble) states. Choosing a flat slicing of dS, one can immediately see that super-Hubble states are `squeezed' due to the curved background, or more specifically, as a consequence of gravity pumping zero-momentum pairs of modes due to a time-dependent mass term in the quadratic Hamiltonian. On the other hand, environment modes are assigned the standard Fock vacuum, since the short-distance behaviour of spacetime is assumed to be that of Minkowski space. The second major complication comes from having a time-dependent interaction parameter which necessitates computing the matrix elements relevant to the problem using time-dependent perturbation theory. This was bypassed in \cite{Brahma:2020zpk} by assuming that the leading-order term is sufficient to capture the relevant physics, the validity of which we shall examine in this work. Finally, another technicality arises regarding the particular configuration of momentum modes that contributes maximally to the integrals of the matrix elements. This was assumed to be the ``squeezed shape'' in \cite{Brahma:2020zpk} due to physical considerations. In this work, we will show more explicitly, through numerical studies, that indeed the momentum integrals can be highly simplified by approximating them with such profiles for momentum triangles, although the shape dominating in this case will be a different one due to the difference in the choice of our interaction term (when compared with what was taken in \cite{Brahma:2020zpk}). In short, we try to generalise the momentum-space calculation carried out in \cite{Balasubramanian:2011wt,Kumar:2017ctm} for a scalar field in Minkowski space to that for one in dS.

With this in mind, one of our main goals is to illustrate that the entanglement entropy between the momentum modes of a scalar field increases rapidly for dS space. This is in line with what was shown to be the case in \cite{Brahma:2020zpk}, under the above-mentioned assumptions, and has to do primarily with the accelerating expansion of the background. In standard Big Bang expansion, where there is no acceleration, momentum modes do not cross the Hubble horizon (rather, they ``re-enter'' the Hubble patch if one assumes inflation to precede such a phase) and we do not expect to see this rapid growth in entanglement entropy due to mode-coupling. Nevertheless, in the concluding section, we will speculate what this growth in entanglement tells us about the nature of dS space itself by comparing it with other well-known measures of entropy and how one can constrain this in the future with bounds from information theory \cite{Gomez:2020xdb,  Bhattacharyya:2020kgu, Aalsma:2020aib,Aalsma:2022swk,Geng:2020kxh,Bedroya:2020rmd, Adhikari:2022oxr}. 

Finally, before delving into the computational aspects of our work, let us also note that momentum space entanglement entropy is not, in general, symmetric with respect to the UV and IR modes, \textit{i.e.,} the result is not invariant under which subsystem we trace out as our environment. This is already true for a bi-partite system in flat space itself, and the demarcation thus plays a role in the final answer. Hence, momentum-space entanglement entropy is not a universal quantity \cite{Kumar:2017ctm}. Strangely, this is one place where the dS space computation fares better than its flat-space cousin -- we have a physically well-motivated reason for choosing the Hubble scale to demarcate the UV from the IR. More specifically, if we have inflation in mind, it makes sense to consider the entropy of the modes which re-enter the horizon and are observed later on while considering the short-wavelength modes as the environment. Thus, even if we know the answer is not independent of the choice of the sub-system partitioning, there is a physical reason for making the choice in this case. Furthermore, as always, the entanglement entropy turns out to be a cut-off dependent quantity. However, once again the relevant cutoff scale for us would be the Planck Mass $\Mpl$, just as the Hubble parameter $H$ demarcates our system from the environment. Our result also depends on the choice of the initial state for the quantum fluctuations, which we assume to be the Bunch-Davies state\footnote{This is a point of contention from the point of view of the trans-Planckian problem of inflation but makes sense for us since we do not want to modify the short-distance behaviour of our theory.} which is a dS-invariant quantity. Furthermore, we assume that there are no superhorizon modes at the beginning of the dS phase, thereby choosing an IR cutoff, and the accelerated expansion creates all the super-Hubble scales of interest. Finally, although we will evaluate the entanglement entropy more rigorously in this work, improving significantly over the approximations made in \cite{Brahma:2020zpk}, there will still be assumptions which we will have to make in our journey (such as assuming that the squeezed states form a complete basis for the super-Hubble modes). We will make these more explicit in the relevant places.

In the main body of the paper, we will lay down the basics of a scalar field theory in dS and assume a cubic interaction term. This is done both since it is the simplest non-linear term that one can consider as well as to remain close to what was done in \cite{Brahma:2020zpk}. Evaluating the perturbative entanglement entropy consists of computing the relevant matrix elements to leading order in perturbation theory. However, we will show how there can be apparent divergences appearing for the time-dependent interaction which one has to deal with appropriately. There are also momentum integrals which are difficult to compute in full generality, where we will use some numerics to show that they peak in a specific ``folded'' limit. This is in line with what is expected from standard cosmological arguments for the bispectrum of such a system. Finally, we will focus on how fast this perturbative quantity is actually growing by comparing it with some large background entropy and end with a speculation regarding what this tells us about the nature of dS in general.

We use natural units throughout this paper, \textit{i.e.} $c = \hbar =1$. In addition, the Planck Mass is denoted by $\Mpl$ and has the same units as the Hubble parameter $H$ while $a$ denotes the scale factor.

\section{Interacting QFT in de Sitter}
\subsection{The Free Theory}
We begin by employing the standard technology of evaluating scalar fields in dS. Since we have inflation as our motivation, we work in the flat slicing of dS, the metric for which is given by
\begin{align}
    d s^2=-a(\eta)^2 \[-d\eta^2+d\bx^2\]\,,
\end{align}
where $\eta =\frac{-1}{aH}$ is the conformal time which runs in the range $-\infty<\eta<0$. We will denote spatial vectors with \textbf{bold} font. 

The Hamiltonian for a free massless scalar is \cite{mukhanov2007introduction}
\begin{align} \label{eq:Free Ham}
\calH_0=\frac{1}{2} \int \frac{d^3 \bk}{(2\pi)^3}  \[k\(c_\bk c_\bk^\dg + c_{-\bk} c_{-\bk}^\dg\) +iaH\( c_\bk c_{-\bk} -c_{-\bk}^\dg c_\bk^\dg\)\],
\end{align}
where $k \equiv \abs{\bk}$, $c_\bk (c_\bk^\dg)$ is the annihilation (creation) operator  and $H^{-1}$ is the characteristic dS (or Hubble) radius. As is well-known, the definition of the vacuum state in dS is not one without ambiguities \cite{Kundu:2011sg}. However, imposing the boundary condition that the mode functions approach Minkowski as $\eta \rightarrow \eta_0 = -\infty$, we can uniquely define a dS-invariant vacuum state known as the  Bunch-Davies (BD) vacuum $\ket{0}_{BD}\equiv\ket{0}$, satisfying the familiar $c_\bk\ket{0}=0$. Sticking to the Heisenberg picture, where the BD state is time-independent, we can work out the time-dependence of the ladder operators through the Bogoliubov transformation
\begin{align}
c_{-\bk}^\dg(\eta)  
 &=e^{i\theta_k} \, \cosh r_k \,c_{-\bk}^\dg(\eta_0) -e^{-i(\theta_k +2\phi_k)} \, \sinh r_k \, c_{\bk}(\eta_0), \label{eq:cq}
\end{align}
where $\theta_k$ and $\phi_k$ represent rotation angles, whereas $r_k$ quantifies the squeezing of the $k^{\rm th}$ mode. These parameters are respectively given by \cite{Martin:2015qta, 2007LNP...738..193M}
\begin{subequations}
\label{SP}
\begin{align}
\theta_{k}(\eta)&=k \eta+\arctan \(\frac{1}{2 k \eta}\),\\
\phi_{k}(\eta)&=\frac{\pi}{4}-\frac{1}{2} \arctan \(\frac{1}{2 k\eta}\), \\
r_{k}(\eta)&=-\arcsinh \(\frac{1}{2k \eta} \)\;.
\end{align}
\end{subequations}
Using this, one can show that modes starting out in the BD vacuum evolve to the squeezed state, on super-Hubble scales, due to the action of the quadratic (free) Hamiltonian \eqref{eq:Free Ham}. The explicit form of the squeezed state can be conveniently written as
\begin{equation}
    \ket{SQ(k,\eta)} \equiv \frac{1}{\cosh{r_k}} \sum_{n=0}^{\infty} e^{-2in \phi_k} \tanh^n r_k \ket{n_{\bf k}, n_{-{\bf k}}}\;,
\end{equation}
where
\begin{equation*}
    \ket{n_{\bf k}, n_{-{\bf k}}} \equiv \frac{1}{n!} (c^{\dagger}_{\bf k}c^{\dagger}_{-{\bf k}})^n \ket{0_{\bf k}, 0_{-{\bf k}}}\;.
\end{equation*}

\subsection{Interacting Theory}
While it is possible to calculate the geometric entanglement entropy inherent to a free theory in de Sitter (see \cite{Maldacena:2012xp, 2014JHEP...07..072K}), we are interested in evaluating the momentum-space entanglement entropy arising from an interaction term. We will closely follow the procedure laid down in \cite{Balasubramanian:2011wt}, developed for deriving the momentum space entanglement entropy from the standard notion of von Neumann entropy. For the sake of clarity, we shall go through some of their arguments here.

The decomposition for a generic perturbed state in a total Hilbert space $\Hilbert=\Hilbert_E \otimes \Hilbert_S$, where $\Hilbert_{E(S)}$ denote environment (system) Hilbert space respectively, in terms of the unperturbed states of both subsystems is given by

\begin{align}\label{Omega}
    \ket{\Omega}= \(\ket{0} +\sum_{n\neq 0} \calA_n\ket{n}\)_E\otimes\(\ket{0}+\sum_{N\neq 0} \calB_N\ket{N}\)_S+\sum_{n,N\neq 0} \(\calC_{n,N}-\calA_n\calB_N\)\ket{n}_E\otimes\ket{N}_S,
\end{align}
where $\calA,\calB,\calC$ are some matrix coefficients. The main assumption here is that the full Hamiltonian for an interacting bipartite QFT can be written as $\calH = \calH_0^{E} + \calH_0^{S} + \calH_I$, where $\calH_0$ is the free Hamiltonian in \eqref{eq:Free Ham} for both system and environment modes while $\calH_I$ denotes the mode-coupling part (we define the specific interaction term for us below in \eqref{eq:Int Ham}). Given the discussion in the previous section, environment (system) modes are in their Fock (squeezed) state, defined by the appropriate number of $c_\bk^\dg(\eta)$ acting on the BD vacuum. From this expression, we can clearly see that without interactions between the two subsystems, the final terms would not exist and  $\ket{\Omega}$ would simply be a separable state, which has zero entanglement. We reiterate that this measure of entanglement is a standard perturbative one in QFT on a curved background and is thus quite distinct from their holographic counterparts \cite{Nishioka:2018khk,Casini:2022rlv}. We are also not measuring the entanglement in the long-range interactions of the Bunch-Davies modes since, essentially, that was a measure of position space entanglement \cite{Maldacena:2012xp}. Nevertheless, perturbative momentum-space entanglement has the potential to carry information corresponding to measurable observables in the CMB \cite{Brahma:2021mng}.


This provides the perfect segue to discuss the first generalisation of calculating entanglement entropy in dS space compared to what was done in \cite{Balasubramanian:2011wt}. For flat space, all the momentum modes, either in the system or environment bands, are taken to be in the perturbed Fock vacuum. However, in our case, this is no longer true. Although all modes start out in the BD vacuum in the far past, when they are well within the horizon, as they exit the horizon, they get squeezed due to the squeezing term in \eqref{eq:Free Ham}. Essentially, this is why we will treat the system modes to be in the squeezed state while the environment modes -- the ones which remain sub-Hubble -- will remain in their vacuum state\footnote{The discussion here is regarding the choice of the quantum state of the perturbations in dS and has nothing to do with treating the background dS itself as a coherent state on top of a Minkowski vacuum\cite{Dvali:2017eba, 2021JHEP...02..104B, 2021ForPh..6900131B}.}.

\subsection{von Neumann Entropy}
Since we are interested in the behaviour of entanglement entropy in dS, we resort to the simplest kind of potential so that we may avoid needless complications. Given the action of a phi-cubic potential
\begin{align}
S_I=\frac{1}{2} \int d^4x \sqrt{-g} \[-\lambda\psi^3 \],
\end{align}
where $\lambda$ is some weak coupling constant that we can tune, we can derive the Hamiltonian in terms of the rescaled field $\psi=a\varphi$ as
\begin{align}
    \calH_I(\eta)= \lambda \, a(\eta)\int{d^3\bx \: \varphi^3\(\eta , \bx \)}.
    \label{eq:Int Ham}
\end{align}
Since our computation is for the perturbative entanglement entropy, in the presence of an interaction term, we first write down the corresponding perturbed vacuum. To see how the perturbed ground state explicitly looks like, we use the fact that for $\calH_I$, we can define the unitary evolution operator
\begin{align}
    \calU_I\(\eta_0,\eta\) \equiv \mathcal{T} e^{-i\int_{\eta_0}^\eta \, d\eta' \, \calH_I (\eta')}, 
\end{align}
$\mathcal{T}$ signifying the time-ordering operator. Perturbation theory then tells us that
\begin{align}
    \ket{\Omega}\approx \ket{0,0}+ \(-i\int_{\eta_0}^\eta \, d\eta' \, \calH_I (\eta')\) \ket{0,0} +\calO\(\lambda^2\), \label{eq:Pert Vac}
\end{align}
where we have used the notation where $\ket{i,j}\equiv \ket{i}_E\otimes\ket{j}_S$. Using perturbation theory, it is easy to rewrite the above expression in terms of matrix elements as in \eqref{Omega}. Finding the von Neumann entropy is straightforward from hereon, simply taking the outer product of $\ket{\Omega}$ and tracing out the environment dofs shows us that, at leading order in $\lambda$, the diagonalised reduced density matrix only depends on one of the matrix elements, namely  $\calC_{n,N}$ \cite{Balasubramanian:2011wt}, so that
\begin{align}
\mathit{S}_{\rm ent}  = -\sum_{n,N \neq 0} \, \abs{\calC_{n,N}}^2 \( \ln{\abs{\calC_{n,N}}^2} -1\) + \calO\(\lambda^3\),
\end{align}
where $\calC_{n,N}$ can be found from standard perturbation theory (by taking the inner product of \eqref{eq:Pert Vac} with $\bra{n,N}$)
\begin{align}
    \calC_{n,N} \approx \bra{n,N}\(-i\int_{\eta_0}^\eta \, d\eta' \, \calH_I (\eta')\) \ket{0,0} +\calO\(\lambda^2\).
\end{align}

\subsection{Momentum Distributions}
Now we arrive to the crux of the argument in \cite{Balasubramanian:2011wt}. Rather than deriving the entanglement entropy in position space \cite{2009JPhA...42X4007C}, we can instead partition the system in terms of Fourier modes with some momentum scale $\mu$ such that now the sums over excited states translates to a sum over momentum modes with at least one below and one above the demarcation scale
\begin{align}
    \sum_{n,N \neq0} \longrightarrow \sum_{\{\bp_i\}\gtrless \mu}.
\end{align}
Enforcing the scale dependence on this demarcation scale requires working with a dimensionful entropy density $\calS=\mathit{S}/V$, which in the infinite volume limit turns the discrete sum into an integral. This entanglement entropy density can now be computed as 
\begin{align} 
\calS &= -\int_{\{\bp_i\}\gtrless \mu} \prod_i^3 d^3 \bp_i \, \[ \abs{\calC_{\{\bp_i\}}}^2 \( \ln{\abs{\calC_{\{\bp_i\}}}^2} -1\) \]+ \calO\(\lambda^3\), \label{eq:S_EE} \\
 \calC_{\{\bp_i\}} &= \bra{\bp_1,\bp_2,\bp_3}\(-i\int_{\eta_0}^\eta \, d\eta' \, \calH_I (\eta')\) \ket{0,0} +\calO\(\lambda^2\), \label{eq:C_pi}
\end{align} 
where in the second expression we have made use of the fact that even with the external modes $n,N$ ranging up to infinitely many excited states in either Hilbert spaces, we know that the matrix elements are exactly zero for all but sets with a number of external states matching the ones created by the interaction Hamiltonian\footnote{Generally only true for cubic and quartic interactions as interactions with higher powers in the field result in counterterms with lower powers and thus allow for a smaller number of external states.}. In our case for the cubic interaction term, this number is three. This means that there are exactly two sets of momenta that we are interested in: set $A$ with 1 sub- and 2 super-Hubble modes, and set $B$ with 1 super- and 2 sub-Hubble modes. To be precise, there are three of each differing only by which label goes where but they are all obviously equivalent, only resulting in some combinatorial numerical factors out in front which we can, as we will with all numerical factors, reabsorb into a redefinition of $\lambda$. Making a particular choice of labelling, we can now write the two sets (setting the beginning of inflation as $a_i=1$)
\begin{align}
    \{\bp_i\}_A &\Rightarrow 
    \left\{
    \begin{array}{ll}
        H<\abs{\bp_1},\abs{\bp_2}< aH\\  
        aH<\abs{\bp_3}<a\Mpl\\
    \end{array}
    \right. ,\nn \\
     \{\bp_j\}_B &\Rightarrow 
    \left\{
    \begin{array}{ll}
        H<\abs{\bp_3}< aH\\  
        aH<\abs{\bp_1},\abs{\bp_2}<a\Mpl\\
    \end{array}
    \right. .\nn 
\end{align}
Given how tedious the calculation will be, we will only show explicitly the entanglement entropy resulting from set $A$ interactions, which we will see later is actually the dominant one, while relegating the other set B to Appendix \ref{sec:Set B}.

\section{Perturbative Momentum-space Entanglement Entropy}
First, notice the mode expansion of the scalar field
\begin{align}\label{eq:modexp}
\varphi \(\eta , \bx \) &= \int{\frac{d^3 \bk}{(2\pi)^3} \frac{1}{\sqrt{2k}}\[c_{\bk}\(\eta\) +c_{-\bk}^\dg \(\eta\)\]e^{i\bk \cdot \bx} } \\
& = \int{\frac{d^3 \bk}{(2\pi)^3} \frac{1}{\sqrt{2k}}\[v_k (\eta) c_{\bk}\(\eta_0\) + v^*_{k} (\eta) c_{-\bk}^\dg \(\eta_0\)\]e^{i\bk \cdot \bx} } \,,
\end{align} 
where the BD mode functions are given by
\begin{equation}
    v_k (\eta) = \frac{e^{-i k \eta}}{\sqrt{2k}} \left(1 - \frac{i}{k \eta} \right)\,.
\end{equation}
Then, we can see that \eqref{eq:Int Ham} becomes an exceedingly large expression of 8 terms all of cubic order in the operators. While not obvious as of yet, it turns out that all the nontrivial terms are of the same form. So rather than working with the whole, we choose a representative of the non-zero terms; $c_{-\bk_1}^\dg c_{-\bk_2}^\dg c_{-\bk_3}^\dg$ and proceed with doing all our calculations with it.

\subsection{Time-Dependence}
We begin by rounding up all the time dependence of \eqref{eq:C_pi} into one expression
\begin{align} \label{eq:I_A}
I_A \equiv  \int_{\eta_0}^\eta d\eta' \, \frac{1}{\eta'} c_{-\bk_1}^\dg(\eta') c_{-\bk_2}^\dg(\eta') c_{-\bk_3}^\dg(\eta'),
\end{align}
where we have used the fact that the scale factor in dS is defined as $a(\eta)=\frac{-1}{H\eta}$. Such that \eqref{eq:C_pi} is now
\begin{align} 
 \calC_{\{\bp_i\}_A} = \bra{\bp_1,\bp_2,\bp_3}\(\frac{-i\lambda}{H}\int_\Delta\frac{I_A}{\sqrt{k_1\,k_2\,k_3}}\) \ket{0,0} +\calO\(\lambda^2\), \label{eq:C_pi eta}
 \end{align}
with the shorthand:  $\int_\Delta \equiv \int \frac{d^3 \bk_1}{(2\pi)^3}\,\frac{d^3 \bk_2}{(2\pi)^3}\,\frac{d^3 \bk_3}{(2\pi)^3}\, (2\pi)^3 \delta^3(\bk_1+\bk_2+\bk_3)$.\\

Substituting in \eqref{eq:cq} into \eqref{eq:I_A}
\begin{align}
    I_A =  c_{-\bk_1}^\dg(\eta_0) c_{-\bk_2}^\dg(\eta_0) c_{-\bk_3}^\dg(\eta_0)  \int_{\eta_0}^\eta d\eta' \, \frac{1}{\eta'}\, e^{i(\theta_{k_1}+\theta_{k_2}+\theta_{k_3})} \cosh r_{k_1} \cosh r_{k_2} \cosh r_{k_3},
\end{align}
where here we have used the property that $c_{\bk}\ket{0}=0$. Evaluating this integral analytically, however, proves to be quite difficult; fortunately, we utilize the reasonable assumption that the dominant behaviour of this integral is still captured in the sub-(super-)Hubble limit of the modes $\left| k\eta\right|\gg 1 \left(\left|k\eta \right|\ll 1 \right)$ \cite{Albrecht:1992kf}, reducing the expression to
\begin{align}
     I_A =  c_{-\bk_1}^\dg(\eta_0)\, c_{-\bk_2}^\dg(\eta_0)\, c_{-\bk_3}^\dg(\eta_0)\, \frac{1}{k_1\,k_2}\, \int_{-\infty }^\eta d\eta'\, \, \frac{1}{\eta'^3}\, e^{iK\eta'}, \label{eq:time int}
\end{align}
with $K\equiv k_1+k_2+k_3$. It is now clear why we chose to analyse one term from \eqref{eq:Int Ham}.

When looking at all the permutations of the ladder operators in \eqref{eq:Int Ham}, only terms with the form of \eqref{eq:time int} survive the right multiplication with the vacuum. In the super-Hubble limit $e^{i\theta_k}$  reduces to $e^{ik\eta}$ + phase, $e^{i \phi_k}$ to a phase and $\sinh r_k \approx \cosh r_k$  to $\frac{1}{2k\eta}$. The outcome is that the terms look identical so when taking the aggregate action of the Hamiltonian on the BD vacuum, we simply get \eqref{eq:time int} with a factor\footnote{Which as we said prior, can be reabsorbed into $\lambda$.} in front. 

To evaluate this integral, we integrate by parts $m$ times such that we have
\begin{align}
    \int_{-\infty }^\eta d\eta' \, \frac{1}{\eta'^3}\, e^{iK\eta'} = \frac{e^{iK\eta}}{2K\eta^3} \sum_{j=0}^{m-1} \(-i\)^{j+1} \, \frac{\(j+2\)!}{\(K\eta\)^j}+\calO\(\frac{1}{\(K\eta\)^{m+1}}\). \label{eq:time sum}
\end{align}
Firstly, note that as long as $K$ is made up of at least one sub-Hubble mode, $\left| K\eta \right| \gg 1 \;\forall\, \eta$, which must be the case to allow for mode-mixing, it is then easy to intuitively see that it is possible to take the  $m\rightarrow\infty$ limit so that we are only left with an exact sum. Doing so, however, clearly results in the sum being divergent. Rather, it is more appropriate to approximate the integral by taking only the first-order term corresponding to $m=1$. Although this seems to be a rather drastic assumption, see Section \ref{sec:RL} for the justification. (In fact, this approximation is nothing but the so-called Riemann-Lebesgue lemma in disguise, as was used in \cite{Brahma:2020zpk} earlier, and will be explained further in the next subsection.) In light of this discussion, the leading order computation gives us
\begin{align}
     I_A =  c_{-\bk_1}^\dg(\eta_0) c_{-\bk_2}^\dg(\eta_0) c_{-\bk_3}^\dg(\eta_0) \frac{-ie^{iK\eta}}{k_1\,k_2 K \eta^3}.
\end{align}
Expanding the states $\bra{\bp_1,\bp_2, \bp_3}$ in terms of the ladder operators, denoting $P$ identically as with $K$, we can rewrite \eqref{eq:C_pi eta} as
\begin{align}
     \calC_{\{\bp_i\}_A} = -\frac{ \lambda}{H\,\eta^5}  \[  \frac{\delta^3\(\bp_1+\bp_2+\bp_3\)}{p_1^2\,p_2^2\,\sqrt{p_1\,p_2\,p_3}\,P}  \] 
\end{align}
where we have implicitly evaluated the momenta integrals using the delta functions obtained from the equal time commutation relations and the standard normalisation that $\braket{0|0}=1$.

Finally, we can write down what the integral expression for the entanglement entropy for set $A$ is given by
\begin{align} \label{eq:S_A}
 \calS_A &=- \int_{\{\bp_i\}\gtrless \mu} \prod_i^3 d^3 \bp_i  \, \[ \abs{\calC_{\{\bp_i\}_A}}^2 \( \ln{\abs{\calC_{\{\bp_i\}_A}}^2} -1\) \]+ \calO\(\lambda^3\), \nn\\
&\approx  -\frac{\lambda^2  \ln{\lambda^2}}{H^2\, \eta^{10}}\, \int_{\{\bp_i\}\gtrless \mu} \prod_i^3 d^3 \bp_i \,\[\frac{\delta^3\(\mathbf{p}_1+\mathbf{p}_2+\mathbf{p}_3 \)}{p_1^5\, p_2^5\, p_3\, P^2} \]  + \calO\(\lambda^2\)
\end{align}
where in the second line we separated the logarithmic term such that we are left only with the leading order. All that is left now is to evaluate these integrals.

\subsection{Hiatus: Breakdown of Perturbation Theory \& Issues of Covariance}\label{sec:RL}
Ideally, to evaluate \eqref{eq:time sum} exactly, we can use the Riemann-Lebesgue lemma \cite{Riemann-Lebesgue}. Simply put, for a function $f(x)$ that is $\mathbb{C}^m$ over an interval $[a,b]$
\begin{align}
    \lim_{p \,\rightarrow \infty}  \int_b^a dx \, e^{ipx} f(x) &= \calO \left(\frac{1}{p}\right), \nn\\
&= \[ e^{ipx} \sum_{j=0}^{m-1}  \frac{(-1)^j}{(ip)^{j+1}} \,f^{(j)}(x) \]_{b}^a +\calO\(\frac{1}{p^{m+1}}\) \label{eq:RL}
\end{align}
where in the second line, the sum comes from integrating by parts, with $f^{(j)}$ being the $j^\text{th}$ derivative. Ostensibly, the first equality makes sense physically because the rapid oscillations of the exponential would tend to cancel out and thus the integral approaches zero. Applying it to \eqref{eq:time sum} is somewhat subtle since we do not have the condition that $K\rightarrow\infty$ but rather $-K\eta>1\; \forall \eta$. This problem is easily overcome since our function is $1/\eta'^3$, so the differentiation always brings down a power of $\eta'$ thus the hope is that we get
\begin{align}
    \int_{-\infty }^\eta d\eta' \, \frac{1}{\eta'^3} e^{iK\eta'}=\calO\(\frac{1}{K\eta}\),
\end{align}
which is well bounded as we would conclude from the lemma. However, the problem we face, as we saw in \eqref{eq:time sum}, is that our series has a vanishing radius of convergence due to the factorial growth. This is another subtlety we can overcome by looking at the special Exponential Integral function \cite{2019arXiv190712373M} defined as
\begin{align}
    \text{Ei}\(z\)\equiv\int_{-\infty}^z\, dt\, \frac{e^t}{t}\,.
\end{align}

If we switch the order of the terms when we integrate by parts, we can get \eqref{eq:time int} in the Exponential Integral form:
\begin{align}
     \int_{-\infty }^\eta d\eta' \, \frac{1}{\eta'^3} e^{iK\eta'}=\frac{-e^{iK\eta}}{2\eta^2} +\frac{-iKe^{iK\eta}}{2\eta}+ \frac{iK}{2\eta}\, \text{Ei}\(iK\eta\).
\end{align}
It is well-known that the series expansion of the Exponential integral is convergent for all complex values of its argument \cite{2019arXiv190712373M}. Indeed, interestingly, the asymptotic expansion for $\abs{K\eta}\gg 1$ turns out to mimic our previous asymptotic series \eqref{eq:time sum} being a factorial divided by a power law. Since the function is bounded, then our integral must be so too, and the apparent divergence must cancel at higher orders. This is not a very uncommon finding in quantum theory where it is well-known that Feynmann diagrams are often accompanied by factorial growth with the perturbative series becoming divergent at some order in the expansion, and therefore requiring a Borel-Ecall\'e resummation for exploring their non-perturbative contribution \cite{2019AnPhy.40967914D, Brahma:2022wdl}. Using similar logic, finding an exact result may need some non-perturbative methods such as resurgence, but it is sufficient for our calculations to rely on the first-order term recovered by the application of the Riemann-Lebesgue lemma.

\subsection{Triangle Integrals}
\begin{figure}[H]
\centering
\begin{tikzpicture}[thick]
\draw (-4,0)coordinate (a) to node[below]{$\bp_3$} (4,0);
\draw (4,0)coordinate(b)  to node[above,rotate=-10]{$\bp_1$} (0,0.5);
\draw  (0,0.5)coordinate(c) to node[above,rotate=10] {$\bp_2$} (-4,0);
\pic [draw=red,angle radius=2cm] {angle = b--a--c};
\node at (-1.6,0.125) {$\omega$};
    \draw [line width=1pt,->-=0.5] (-4,0)-- (4,0);
    \draw [line width=1pt,->-=0.5] (4,0)-- (0,0.5);
    \draw [line width=1pt,->-=0.5] (0,0.5)-- (-4,0);
\end{tikzpicture}
\caption{Geometric representation of an interaction between three Fourier modes.}\label{fig:triangle A}
\end{figure}
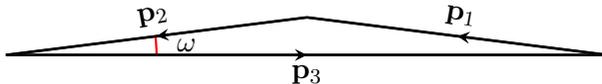

Let us return to \eqref{eq:S_A} and try to compute the set of momentum integrals. The first obvious step is to kill one of the integrals using the delta function. There is, however, a subtlety here; this is only possible since the peak of the function is within the range of the integral. Choosing, rather arbitrarily, to kill the $d^3\bp_1$ integral, we can get a relationship between the energies $p_1= \sqrt{p_2^2+p_3^2-2\,p_2\,p_3 \cos{\omega}}$, where $\omega$ is the angle between $\bp_2$ and $\bp_3$. In Fourier space, this closes a triangle made up of the three momenta and this can be seen in Figure \ref{fig:triangle A}.

Orienting the triangle such that $\bp_2$ is parallel to the z-axis, we can see that $d^3\bp_2=4\pi p_2^2 \,dp_2$ and $d^3\bp_3=2\pi p_3^2 \,dp_3 d\left(-\cos \omega\right)$. We can now write \eqref{eq:S_A} with the appropriate limits on the energies
\begin{align}
     \calS_A \approx -\frac{\lambda^2  \ln{\lambda^2}}{H^2\, \eta^{10}}\, \int_H^{a H} dp_2 \int_{aH}^{a \Mpl} dp_3\int_{\alpha_a(p_2,p_3)}^{\alpha_b(p_2,p_3)} d\(-\cos{\omega}\) \[\frac{p_3}{P^2\, p_2^3\, \(p_2^2+p_3^2-2\,p_2\,p_3 \cos{\omega}\)^{5/2}} \]\label{eq:S_A angle}
\end{align}
where we have introduced the variables $\alpha_a,\alpha_b$ as angular limits which are dependent on both momenta since the angle determines the shape of the triangle. Knowing the range of $(\bp_3-\bp_2)^2=\bp_1^2$, we can write the range of $-\cos\omega$ such that
\begin{align}
   \alpha_a \equiv \frac{1}{2p_2p_3}\[( H)^2-p_2^2-p_3^2\] < &-\cos \omega <  \frac{1}{2p_2p_3}\[(a H)^2-p_2^2-p_3^2\]\equiv \alpha_b. \label{eq:angle in mom terms}
\end{align}
The limits for the momentum integrals can be understood as follows. Since we have chosen the Planck mass as the UV cut-off for our theory, it is no surprise that the resulting entanglement entropy will then depend on this UV scale. The comoving Hubble scale $aH$ is what demarcates system from environment modes and thus is the lower limit for the UV modes and the upper limit for the IR ones. Finally, we postulate that there were no superhorizon scales at the beginning of the dS phase (where we have set $a_i = 1$), \textit{i.e.} all the super-Hubble modes were created by the dS expansion.

Once more, we are faced with an integral that is not trivial to solve. As expected, even if the limits of the magnitude of the remaining momenta (after using the delta function) are straightforward to evaluate, this is not the case for the angular integral which has the limits as a function of the momenta. However, a little bit of reflection shows that the integral is dominated by a term that saturates the angle in the configuration when the Triangle \ref{fig:triangle A} is \textit{folded}, \textit{i.e.} $\omega\approx 0$ (See \ref{sec:Angle A}). Being left with integrals with the momenta in the denominator, we easily see that the IR limit dominates as we approach late times $a\gg 1$ and so the entanglement entropy, at leading order, is given by 
\begin{align}
    \calS_A \approx -\lambda^2 \ln\lambda^2 a^{10} H. \label{eq:S_EE A}
\end{align}
The first thing to notice is that the UV cutoff does not show up in this answer since the IR limit dominates the integral and this is a direct consequence of squeezing. In other words, since our interaction is of the $\phi^3$ form, and does not have any derivative interactions as is typically the case for gravity \cite{Brahma:2020zpk}, the folded shape dominates in which case we have two super-Hubble and one sub-Hubble mode. This results in an extra factor of the squeezing (as compared to the so-called ``squeezed'' limit in \cite{Brahma:2020zpk}) and the IR limit dominating the integral.

Performing a similar analysis for set $B$, we find that it is sub-dominant to set $A$, explaining why we left it out of the main text and relegated it to the appendices, namely,
\begin{align}
    \calS_B \approx -\lambda^2 \ln\lambda^2 a^{6} H.
\end{align}

\section{Conclusions}
Let us first convert our result to something a bit more physically meaningful -- this would be the entanglement entropy per unit physical volume, which is given by 
\begin{align}
    s_{\rm EE} \sim \lambda^2 a^{7} H\,, \label{eq:S_EE Final}
\end{align}
where we have ignored small logarithmic corrections to focus on the leading order contribution. A quick comparison with the entanglement entropy (per unit physical volume) of cosmological perturbations shows that \cite{Brahma:2020zpk} the growth here is even faster than in that case. The reason is the same as explained above. The interaction term chosen here is simply devoid of any spatial derivatives and this leads to a stronger dependence on the IR modes, and hence a faster growth of the entanglement entropy. 

Although we computed the entanglement entropy between sub and super-Hubble modes for a spectator scalar field in (the flat slicing of) dS, it does have an important physical application. One would expect a similar contribution for the tensor modes in inflation since the leading order term for the cubic non-Gaussianities for purely primordial gravitational waves is free of any slow roll parameter, and indeed has a term devoid of any derivatives (see {\it e.g.} \cite{Brahma:2022yxu, Gong:2019yyz}). Thus, one can expect a similar behaviour for the growth of entanglement entropy for inflationary tensor modes. Thus, any conclusions which could be drawn from the growth of this perturbative entanglement entropy of a scalar field (due to the background evolution) during inflation can also be drawn from the tensor entanglement entropy. However, it remains to be seen what physical implications can we actually draw from such a momentum space computation, especially with respect to observations.

\subsection{Interpretation}
One way to draw some physical consequences for this would be to compare this entanglement entropy with the thermal entropy during reheating. This is reiterating the argument put forward in \cite{Brahma:2020zpk} which can be stated as follows. If we assume that all of this entanglement entropy gets converted into a thermal entropy, how large can this entanglement entropy be so as to not become greater that the reheating entropy? This was answered in \cite{Brahma:2020zpk} for scalar perturbations and we can simply reproduce the calculation here. In other words, the time scale on which the entanglement entropy becomes comparable to the thermal entropy of reheating ($s_{\rm th}\sim  \Mpl^{3/2} H^{3/2}$) is given by
\begin{eqnarray}
 N \simeq \frac{1}{7} \ln \left(\frac{\Mpl^{3/2} H^{1/2}}{\lambda^2}\right)\,,
\end{eqnarray}
where $N\equiv (a/a_i)$. Qualitatively, this gives the same bound as in \cite{Brahma:2020zpk}, \textit{i.e.} the entanglement entropy around the scrambling time of dS becomes large enough to account for all of the reheating entropy\footnote{The astute reader will notice that the above time scale is larger then the scrambling time ($N<\ln{\Mpl/H}$) but it is not by much since the relevant quantity appears inside a logarithm. Of course, if $\phi$ is a spectator field, then $\lambda \ll H$ and the above quantity is larger than the scrambling time and is equal to it in the limit that the cubic potential is responsible for the accelerating expansion $\lambda \sim H$. However, our main point is that even for a perturbative potential $\lambda/H \ll 1$, this time scale is not a very large one since it appears inside a logarithm and demonstrates the rapid growth of the entanglement entropy.}

Of course, it is entirely possible that the momentum-space entanglement entropy is not converted into a thermal entropy and then the upper bound does not apply in this case. More so, in our original computation, we simply assumed a spectator scalar field in pure dS and without invoking the analogy to primordial tensor modes, a comparison with the reheating entropy is not applicable.

In its current state, the only conclusion about the EE \eqref{eq:S_EE A} we can make is that grows quite fast. However, to gain some insight into how fast is its rate of growth, we can compare it to other known entropy results for dS. For instance, one might compare this with the well-known Gibbons-Hawking (GH) entropy \cite{Gibbons:1977mu}. However, note that we cannot quite apply a Bousso bound \cite{Bousso:1999xy} to our computation since ours is an entanglement entropy between momentum modes of a scalar field which live everywhere and is not restricted within a static patch of dS. We are simply comparing this to the GH entropy to see how fast can this tiny perturbative computation become as large that the former quantity.

Since we are comparing the entanglement entropy per unit physical volume, we must divide the total GH entropy $\calS_{\rm GH}=\frac{\Mpl^2}{H^2}$ with the dS physical volume $V_{dS}=(H)^{-3}$ so that now
\begin{align}
    s_{\rm EE}\sim \lambda^2 a^{7} H  \leq \Mpl^2 H \simeq s_{\rm GH} \label{eq:GH compare}
\end{align}
where we have ignored the logarithmic term in the coupling, as before, as it is negligible compared to the quadratic term. Recalling that we had set the initial value of the scale factor $a_i=1$, this shows that around the time
\begin{align}
    N\simeq \frac{2}{7}\ln \frac{\Mpl}{\lambda}\,,
\end{align}
where this number of $e$-foldings denotes when the perturbative entanglement entropy due to the cubic interaction term becomes as large as the Gibbons-Hawking entropy. However, as mentioned above, one should not look at this as a bound for the number of e-foldings allowed for the dS phase to exist. Rather, this is just to give a measure of how fast the entanglement entropy is growing in this system.

If we cannot use the reheating entropy or the GH entropy to put a bound on the lifetime of the dS phase, how should we interpret our result? At the very least, this computation shows us the importance of time-scale $\ln (\Mpl/\lambda)$, after which we should not trust our perturbative calculation.  Note that although $\Mpl/\lambda$ is much longer than the scrambling time of dS, given by $\Mpl/H$, it is still a very small amount since its logarithm is the one which appears as the relevant timescale. This sets a limit on how long such a perturbative treatment is under control for a QFT in dS space due to the secular growth in the entanglement entropy. Since our computation of the entanglement entropy is in momentum space, it is difficult to put any direct bounds on it from some physical reasoning. However, more ambitiously, we might be able to put a bound on the growth rate of the entanglement entropy coming from upper limits on its velocity from quantum information theory \cite{Hartman:2015apr,Couch:2019zni,Folkestad:2022huc}. This might open up an interesting avenue to constrain perturbative QFT in dS space.

In summary, in this work we have looked at a perturbative scalar QFT in dS space and computed the entanglement entropy, in momentum-space, between the sub- and super-Hubble modes of the scalar field. Being in momentum space, the entire entanglement results from the perturbative non-linearity and our main result is to demonstrate how fast this quantity grows due to IR effects of dS space. At the very least, our result shows how long one can trust perturbative results, especially in relation to entanglement between fields in dS space, before secular effects take over and one needs to employ some late-time (presumably, non-perturbative) resummations to deal with them. A final caveat to keep in mind is that we have used the planar slicing of dS, which is relevant especially for discussing cosmological accelerating spacetimes, and yet one must compute similar quantities in other slicings of dS to better understand the crucial features exhibited by the background expansion on such entanglement. We leave this and other intriguing issues mentioned above for future work.

\section*{Acknowledgements}
SB thanks Alek Bedroya and Robert Brandenberger for interesting discussions.

SB is supported in part by the Higgs Fellowship and by the STFC Consolidated Grant “Particle Physics at the Higgs Centre”. For the purpose of open access, the authors have applied a Creative Commons Attribution (CC BY) license to any Author Accepted Manuscript version arising from this submission.

\begin{appendices}
\section{Entanglement Entropy for Set B} \label{sec:Set B}
Starting from the equivalent expression of \eqref{eq:C_pi eta} for set B
\begin{align} 
 \calC_{\{\bp_i\}_B} = \bra{\bp_1,\bp_2,\bp_3}\(\frac{-i\lambda}{H}\int_\Delta\frac{I_B}{\sqrt{k_1\,k_2\,k_3}}\) \ket{0,0} +\calO\(\lambda^2\), \label{eq:C_pi eta B}\\
 I_B =  c_{-\bk_1}^\dg(\eta_0) c_{-\bk_2}^\dg(\eta_0) c_{-\bk_3}^\dg(\eta_0) \frac{1}{k_3} \int_{-\infty }^\eta d\eta' \, \frac{1}{\eta'^2} e^{iK\eta'},
 \end{align}
 where this time we have taken the sub- and super-Hubble limits for $k_1,k_2$ and $k_3$, respectively.\\

 Integrating by parts and taking the leading order, we get
 \begin{align}
     I_B= c_{-\bk_1}^\dg(\eta_0) c_{-\bk_2}^\dg(\eta_0) c_{-\bk_3}^\dg(\eta_0) \frac{-ie^{iK\eta}}{k_3 K \eta^2} 
 \end{align}
Performing a similar analysis as before, we see that \eqref{eq:C_pi eta B} is now
\begin{align}
     \calC_{\{\bp_i\}_B}= -\frac{ \lambda}{H\,\eta^3}  \[  \frac{\delta^3\(\bp_1+\bp_2+\bp_3\)}{p_3^2\,\sqrt{p_1\,p_2\,p_3}\,P}  \].
\end{align}
which implies that the expression for the entanglement entropy for this configuration is given by
\begin{align}
    \calS_B \approx  -\frac{\lambda^2  \ln{\lambda^2}}{H^2\, \eta^{6}}\, \int_{\{\bp_i\}\gtrless \mu} \prod_i^3 d^3 \bp_i \,\[\frac{\delta^3\(\mathbf{p}_1+\mathbf{p}_2+\mathbf{p}_3 \)}{p_1\, p_2\, p_3^5\, P^2} \]  + \calO\(\lambda^2\)
\end{align}

\begin{figure}[H]
\centering
\begin{tikzpicture}[thick]
\draw (-4,0)coordinate (a) to node[below]{$\bp_1$} (4,0);
\draw (4,0)coordinate(b)  to node[above,rotate=-10]{$\bp_2$} (-4,0.5)coordinate (c);
\draw  (-4,0.5) to node[left] {$\bp_3$} (-4,0);
\pic [draw=red,angle radius=3mm] {angle = a--c--b};
\node at (-3.6,0.25) {$\omega$};
    \draw [line width=1pt,->-=0.5] (-4,0)-- (4,0);
    \draw [line width=1pt,->-=0.5] (4,0)-- (-4,0.5);
    \draw [line width=1pt,->-=0.65] (-4,0.5)-- (-4,0);
\end{tikzpicture}
\caption{Geometric representation of an interaction between three Fourier modes.}\label{fig:triangle B}
\end{figure}
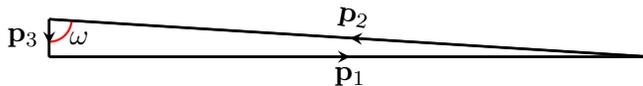

 Killing the $d^3\bp_1$ integral with the delta function and expanding the volume elements, we find
\begin{align}
     \calS_B \approx -\frac{\lambda^2  \ln{\lambda^2}}{H^2\, \eta^{6}}\, \int_{a H}^{a \Mpl} dp_2 \int_{ H}^{a H} dp_3\int_{\beta_a(p_1,p_3)}^{\beta_b(p_1,p_3)} d\(-\cos{\omega}\) \[\frac{p_2}{P^2\, p_3^3\sqrt{p_2^2+p_3^2-2\,p_1\,p_3 \cos{\omega}}}\] \label{eq:S_B angle}
\end{align}
where now the angle ranges, in this case, are determined by the full range of $p_1$, \textit{i.e.}
\begin{align}
   \beta_a= \frac{1}{2p_2p_3}\[(a H)^2-p_2^2-p_3^2\] < &-\cos \omega <  \frac{1}{2p_2p_3}\[(a \Mpl)^2-p_2^2-p_3^2\]=\beta_b. \label{eq:angle B in mom terms}
\end{align}
From Appendix \ref{sec:Angle B} we know that the angular integral peaks when $\omega \approx \frac{\pi}{2}$ in the so-called ``squeezed'' limit. Performing the momenta integrals, we again see that the IR limits dominate, as perhaps expected, with the result that
\begin{align}
    \calS_B \approx -\lambda\ln\lambda^2 a^{6} H.
\end{align}

\section{Dominant Triangle Shapes for the Momenta Integrals} 
\subsection{Angles in Set A} \label{sec:Angle A}
Starting from \eqref{eq:S_A angle}, we try to simplify the expression by figuring out at which $\omega$ the integral dominates and thus precluding the need to evaluate the full integral. To do that, we need to perform the angular integral, so that we now have
\begin{align}
     \calS_A \approx -\frac{\lambda^2  \ln{\lambda^2}}{H^2\, \eta^{10}}\, \int_H^{a H} dp_2 \int_{aH}^{a\Mpl} dp_3 \, F\(p_2,p_3\) \label{eq:S_A F}.
\end{align}
Showing the resulting function is not very illuminating so we have opted against explicitly writing it down. Rather, we are interested in finding at which point $F\left(p_2,p_3\right)$ gives its maximum contribution to the integral by finding the shape of the angle which saturates at that point.
\begin{figure}[H]
    \centering
    \includegraphics[width=10cm]{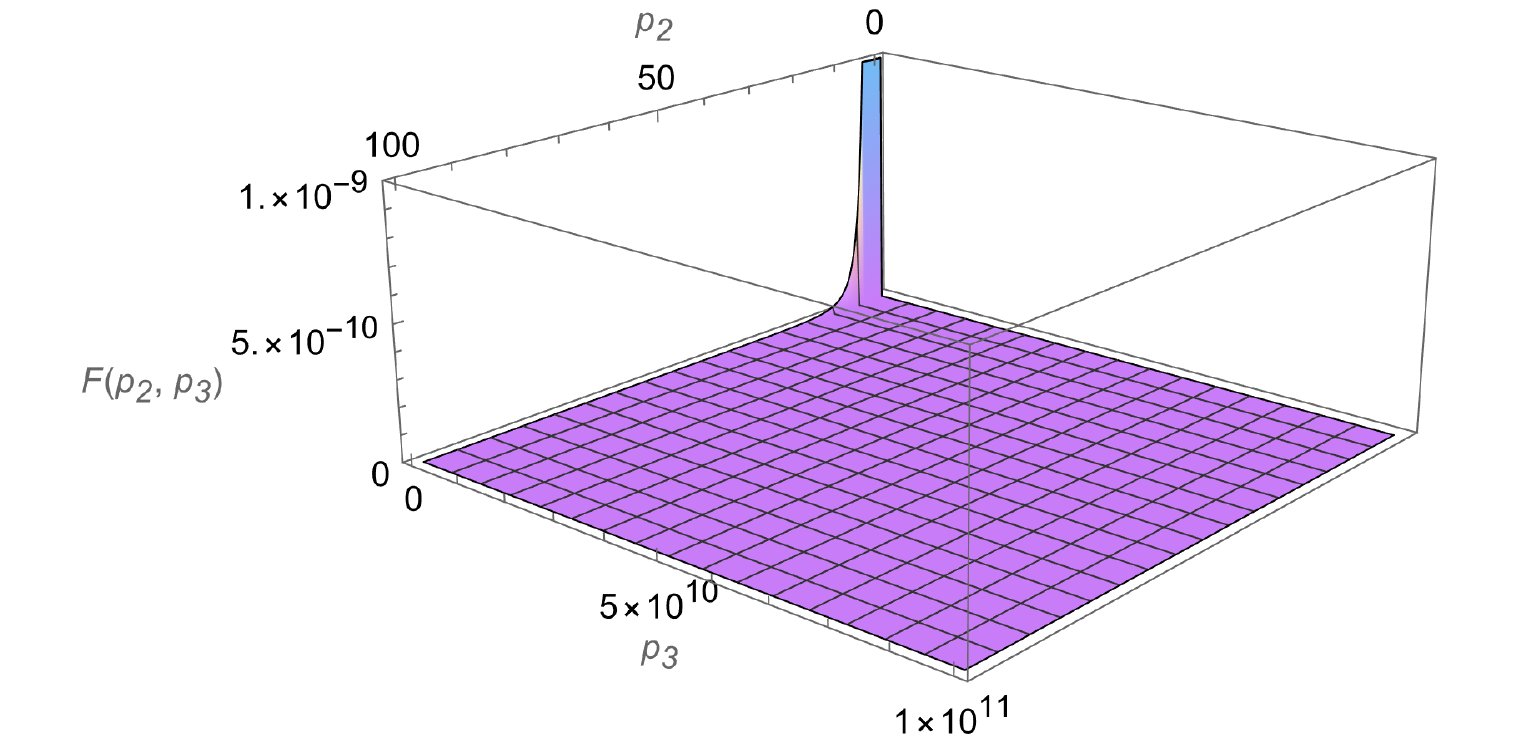}
    \caption{A plot showing the dependence of $F\left(p_2,p_3\right)$, in \eqref{eq:S_A F}, on it arguments. Since we are interested in late-time behaviour, we have chosen reasonable values for the scale factor, Hubble parameter and Planck mass, at $a=100$, $H=1$ and $\Mpl=10^9$, respectively.}
    \label{fig:Folded plot}
\end{figure}
We can see from Figure \ref{fig:Folded plot} that the function peaks in the limits $\left(p_2,p_3\right)\rightarrow\left(H,aH\right)$. These values severely limit the shape of the Triangle \ref{fig:triangle A}, taking into account the range of $p_1$ we can see that it must be folded. This is sufficient justification for saying that if we are interested in the dominant part of \eqref{eq:S_A angle}, we may bypass the angular integral and simply take the corresponding angle to be that of the folded shape $\omega \approx 0$.

\subsection{Angles in Set B} \label{sec:Angle B}
\begin{figure}[H]
    \centering
    \includegraphics[width=10cm]{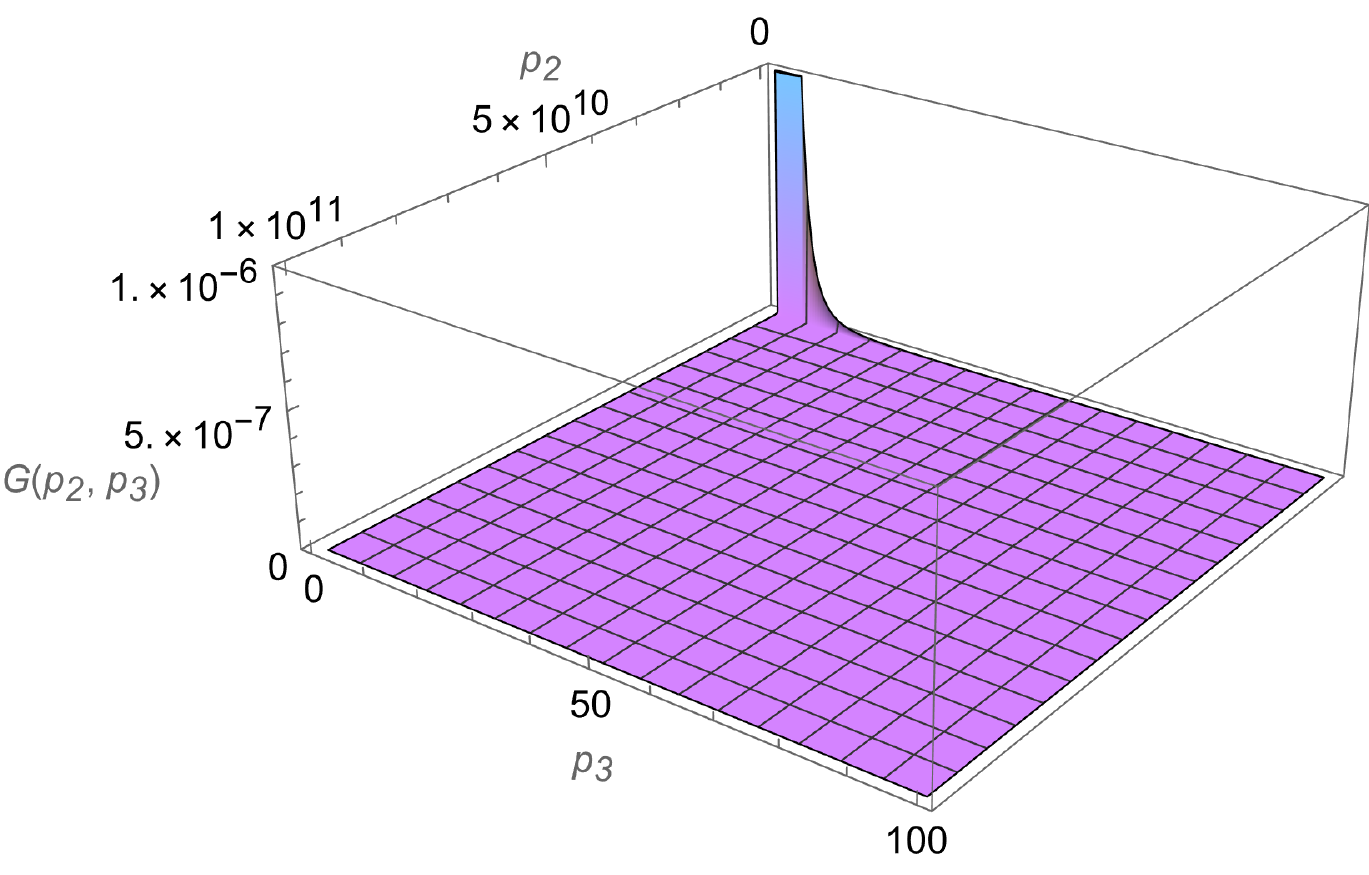}
    \caption{A plot showing the dependence of $G\left(p_2,p_3\right)$, in \eqref{eq:S_B G}, on it arguments. Since we are interested in late-time behaviour, we have chosen reasonable values for the scale factor, Hubble parameter and Planck mass, at $a=100$, $H=1$ and $\Mpl=10^9$, respectively.} \label{fig:Squeezed plot}
\end{figure}

Performing the angular integral in \eqref{eq:S_B angle}
\begin{align}
     \calS_B \approx -\frac{\lambda^2  \ln{\lambda^2}}{H^2\, \eta^{6}}\, \int_{a H}^{a \Mpl} dp_2 \int_{H}^{aH} dp_3 \, G\(p_2,p_3\) \label{eq:S_B G}
\end{align}
we similarly express the result in terms of a function $G\left(p_2,p_3\right)$ which we want to plot the behaviour of. Plotting the function $G\left(p_2,p_3\right)$ against its argument, we can see, from Figure \ref{fig:Squeezed plot}, that it peaks, similarly, in the IR limits of the momenta. Once more, taking into account the range of $p_1$, the configuration corresponding to those limits is when the angle in Triangle \ref{fig:triangle B} is $\omega \approx \frac{\pi}{2}$ which is the squeezed shape in the standard cosmological vernacular.

\end{appendices}



\printbibliography

\end{document}